\documentclass[12pt,notitlepage]{article}
\usepackage{amsmath,amssymb,amsthm}
\usepackage{cite}
\begin{document}
\begin{center}{\Large \bf Integrable Rosochatius deformations of higher-order constrained flows and the soliton
 hierarchy with self-consistent sources}
\end{center}
\begin{center}
{ Yuqin Yao\footnote{yqyao@math.tsinghua.edu.cn} and   Yunbo
Zeng\footnote{yzeng@math.tsinghua.edu.cn}}
\end{center}

\begin{center}{{\small \it Department of Mathematics,
Tsinghua University, Beijing 100084 , PR China}}
\end{center}
\vskip 12pt { \small\noindent\bf Abstract}
 {We propose a systematic method to generalize the integrable Rosochatius deformations for finite dimensional
 integrable Hamiltonian systems to  integrable Rosochatius deformations for infinite dimensional integrable
 equations. Infinite number of the integrable Rosochatius
 deformed higher-order constrained flows of some soliton hierarchies, which
 includes the generalized integrable H$\acute{e}$non-Heiles system,
and the integrable Rosochatius deformations of
 the KdV hierarchy with self-consistent sources, of the AKNS hierarchy with self-consistent sources
  and of the
 mKdV hierarchy with self-consistent sources
 as well as their
 Lax representations are presented. }\\

\section{Introduction}
 Rosochatius found that it would still keep the
integrability to add a potential of the sum of inverse squares of
the coordinates to that of the Neumann system\cite{1,2}. The
deformed system is called Neumann-Rosochatius system. Wojciechowski
obtained an analogy system for the Garnier system as a stationary
KdV flow in 1985\cite{3,4}. In 1997, Kubo et al.\cite{7} constructed
the analogy system for the Jacobi system\cite{5} and the geodesic
flow equation on the ellipsoid based upon the Deift technique and a
theorem that the Gauss map transforms the Neumann system to the
Jacobi system\cite{moser,knorrer}. All of these systems have the
same character that they are integrable Hamiltonian systems
containing $N$ arbitrary parameters and the original finite
dimensional integrable Hamiltonian systems are recovered when all
these parameters vanish. In fact, these systems are a sort of
integrable deformations of the corresponding integrable Hamiltonian
systems, which are called integrable Rosochatius deformation. The
resulting systems are called the Rosochatius-type integrable
systems. These systems have important physical applications. For
examples, Neumann-Rosochatius system can be used to describe the
dynamics of rotating closed string solutions in $AdS_{5}\times
S^{5}$ and the membranes on $AdS_{4}\times S^{7}$\cite{8}-\cite{13}.
The Garnier-Rosochatius system  can be used to solve the
multicomponent coupled nonlinear Schr$\ddot{o}$dinger
equation\cite{3,4,14}. It is not difficult to see that each of the
above system has its own origin. In Ref.\cite{zhou}, Zhou generalize
the Rosochatius method to study the integrable Rosochatius
deformations of some explicit constrained flows of soliton
equations.

However, so far the Rosochatius deformations are limited to few
finite dimensional integrable Hamiltonian systems(FDIHSs). It is
natural to ask whether there exist integrable Rosochatius
deformations for infinite dimensional integrable equations. The main
purpose of this paper is to generalize the Rosochatius deformation
from FDIHSs to infinite dimensional integrable equations. We will
investigate the integrable Rosochatius deformations firstly for
infinite number of higher-order constrained flows of some soliton
equations, then for some soliton hierarchies with self-consistent
sources.

In recent years the
 constrained flows of soliton equations obtained from the symmetry
reduction of soliton equations, which can be transformed into a
FDIHS, attracted a lot of
attention\cite{constrainedflow1}-\cite{constrainedflow12}. Many
well-known FDIHSs are recovered by means of constrained flows.
Furthermore, soliton equation can be factorized into two commuting
constrained flows, which provides an effective way to solve the
soliton equations through solving the constrained flows.

To make the paper self-contained, we first briefly recall the
 higher-order constrained flows of the soliton hierarchy and
 the soliton hierarchy with
 self-consistent sources.
Consider a hierarchy of soliton equation which can be formulated as
an infinite dimensional Hamiltonian systems
\begin{equation}
\label{eqns:HS} u_{t_{n}}=J \frac{\delta H_{n}}{\delta u}.
\end{equation}
The auxiliary linear problems associated with (\ref{eqns:HS}) are
given by
\begin{subequations}
\label{eqns:linearprob}
  \begin{align}
    &\phi_{x}=U(\lambda,u)\phi,~\phi=(\phi_{1},\phi_{2})^{T}
\\
    &\phi_{t_{n}}=V^{(n)}(\lambda,u)\phi.
  \end{align}
\end{subequations}
Then the higher-order constrained flows of (\ref{eqns:HS}) consist
of the equations obtained from the spectral problem (2a) for $N$
distinct $\lambda_{j}$ and the restriction of the variational
derivatives for the conserved quantities $H_{n}$ and $\lambda_{j}$
\cite{zeng2}
\begin{subequations}
\label{eqns:higherocf}
  \begin{align}
    &J[\frac{\delta H_{n}}{\delta u}+
\sum\limits_{j=1}^{N}\frac{\delta\lambda_{j}}{\delta u}]=0,
\\
    &\phi_{j,x}=U(\lambda_{j},u)\phi_{j},~\phi_{j}=(\phi_{1j},\phi_{2j})^{T},~j=1,2,\cdots,N.
  \end{align}
\end{subequations}
By introduction the so called Jacobi-Ostrogradsky
coordinates\cite{hh}, (\ref{eqns:higherocf}) can be transformed into
a FDIHS. The Lax representation of (\ref{eqns:higherocf}) can be
deduced from the adjoint representation of (\ref{eqns:linearprob})
\cite{constrainedflow4}
\begin{equation}
\label{eqns:adjointr}
 N^{(n)}_{x}=[U,N^{(n)}].
\end{equation}
Mainly, $N^{(n)}$ has the following forms
$$
N^{(n)}(\lambda)=V^{(n)}(\lambda)+N_{0}=\left(\begin{array}{cc}
 A(\lambda)& B(\lambda)\\
C(\lambda)&-A(\lambda)\\
 \end{array}\right),
$$
where $$N_{0}=
\sum\limits_{j=1}^{N}\frac{1}{\lambda-\lambda_{j}}\left(\begin{array}{cc}
 \phi_{1j} \phi_{2j}& - \phi_{1j}^{2}\\
 \phi_{2j}^{2}&-\phi_{1j} \phi_{2j}\\
 \end{array}\right),~~~
or ~~~
N_{0}=\sum\limits_{j=1}^{N}\frac{1}{\lambda^{2}-\lambda_{j}^{2}}
 \left(\begin{array}{cc}
 \lambda\phi_{1j} \phi_{2j}& - \lambda_{j}\phi_{1j}^{2}\\
\lambda_{j}\phi_{2j}^{2}&-\lambda\phi_{1j} \phi_{2j}\\
\end{array}\right).
$$
The soliton hierarchy  with self-consistent sources is defined by
\cite{constrainedflow5,constrainedflow6}
\begin{subequations}
\label{eqns:sescs}
  \begin{align}
    &u_{t_{n}}=J[\frac{\delta H_{n}}{\delta u}+
\sum\limits_{j=1}^{N}\frac{\delta\lambda_{j}}{\delta u}],
\\
    &\phi_{j,x}=U(\lambda_{j},u)\phi_{j},~j=1,2,\cdots,N.
  \end{align}
\end{subequations}
Since the higher-order constrained flows (\ref{eqns:higherocf}) are
just the stationary equations of (\ref{eqns:sescs}), the
zero-curvature representation for (\ref{eqns:sescs}) can be induced
from (\ref{eqns:adjointr}) as follows
\begin{equation}
\label{eqns:zeror}
 U_{t_{n}}-N^{(n)}_{x}+[U,N^{(n)}]=0
\end{equation}
which implies that (\ref{eqns:sescs}) is Lax integrable. In fact,
(\ref{eqns:sescs}) can also be formulated as an infinite-dimensional
integrable Hamiltonian system with t-type Hamiltonian operator by
taking $t$ as the 'spatial' variable and $x$ as the evolution
parameter as well as introducing the Jacobi-Ostrogradsky
coordinates\cite{blaszak,yzeng}. Then we can find the Poisson
bracket defined by the t-type Hamiltonian operator, and the
conserved density when considering $x$ as the evolution parameter.
The t-type bi-Hamiltonian description for KdV equation with
self-consistent sources and for Jaulent-Miodek equation with
self-consistent sources were presented in \cite{blaszak} and
\cite{yzeng}, respectively.

The soliton equation with self-consistent sources have important
physical application, for example, the KdV equation with
self-consistent sources  describes the interaction of long and short
capillary-gravity wave\cite{constrainedflow14}-\cite{vk2}.

 In this paper, in the same way\cite{zhou}, we construct the
Rosochatius deformation $\tilde{N}^{(n)}$ of Lax matrix $N^{(n)}$ by
replacing $\phi_{2j}^{2}$ in the matrix $N_{0}$ with
$\phi_{2j}^{2}+\frac{\mu_{j}}{\phi_{1j}^{2}}$, namely the entries of
$\tilde{N}^{(n)}$ are given by
$$
\tilde{A}(\lambda)=A(\lambda),~\tilde{B}(\lambda)=B(\lambda),
\tilde{C}(\lambda)=C(\lambda)+\sum\limits_{j=1}^{N}\frac{\mu_{j}}{(\lambda-\lambda_{j})\phi_{_{1j}}^{2}},
$$
or
\begin{equation}
\label{eqns:7}\tilde{C}(\lambda)=C(\lambda)+\sum\limits_{j=1}^{N}\frac{\lambda_{j}\mu_{j}}
{(\lambda^{2}-\lambda_{j}^{2})\phi_{_{1j}}^{2}}\end{equation} Then
Lax representation (\ref{eqns:adjointr}) with $N^{(n)}$ replaced by
$\tilde{N}^{(n)}$ gives rise to the Rosochatius deformations of
(\ref{eqns:higherocf}).
 The fact that such a substitute keeps
the relations of the Poisson brackets of $A(\lambda),~B(\lambda)$
and $C(\lambda)$  guarantees the integrability of Rosochatius
deformations of (\ref{eqns:higherocf}). In this way we can obtain
infinite number of integrable Rosochatius deformed FDIHSs
constructed from the Rosochatius deformed higher-order constrained
flows of KdV hierarchy, AKNS hierarchy and mKdV hierarchy,
respectively. Among these Rosochatius deformed FDIHSs, it needs to
point out that the Rosochatius deformation of the first higher-order
constrained flow of KdV hierarchy contains the well-known
generalized integrable H$\acute{e}$non-Heiles system, and can be
regarded  as the integrable multidimensional extension of
H$\acute{e}$non-Heiles system.
 Then, based on the Rosochatius deformed higher-order constrained flows, the
Rosochatius deformations of the soliton hierarchy with
self-consistent sources (RDSHSCS) can be constructed through
(\ref{eqns:zeror})  with $N^{(n)}$ replaced by $\tilde{N}^{(n)}$.
The integrability of the RDSHSCS can be explained by the fact that
the RDSHSCS possesses the zero-curvature representation
(\ref{eqns:zeror})  with $N^{(n)}$ replaced by $\tilde{N}^{(n)}$ and
its stationary reduction  is integrable Rosochatius deformations of
(\ref{eqns:higherocf}). In these way, we construct Rosochatius
deformations of KdV hierarchy with self-consistent sources
(RDKdVHSCS), of AKNS hierarchy with self-consistent sources
(RDAKNSHSCS) and of mKdV hierarchy with self-consistent sources
(RDmKdVHSCS), as well as their zero-curvature representations. We
notice that there are two kinds of RDSHSCS. For RDKdVHSCS and
RDmKdVHSCS, the Rosochatius formed terms only appear in (5b).
However for RDAKNSHSCS Rosochatius deformed terms occur in both (5a)
and (5b).

The structure of the paper is as follows. In Sec.2, we present the
Rosochatius deformation of the higher-order constrained flows of KdV
hierarchy and RDKdVHSCS as well as  their Lax representations. In
Sec.3, we obtain the Rosochatius deformation of  higher-order
constrained flows of AKNS hierarchy and RDAKNSHSCS as well as their
Lax representations. In Sec.4, we obtain the Rosochatius deformation
of higher-order constrained flows of mKdV hierarchy and RDmKdVHSCS
and their Lax representations. In Sec.5, a conclusion is made.

\section{\bf The Rosochatius deformed KdV hierarchy with self-consistent sources}

 Consider the Schr$\ddot{o}$dinger equation \cite{constrainedflow13}
\begin{equation}
\label{eqns:schrodinger} \phi_{1xx}+(\lambda+u)\phi_{1}=0,
\end{equation} which can be written in the matrix form
\begin{equation}
\label{eqns:matrixform} \left(\begin{array}{c}
 \phi_{1}\\
\phi_{2}\\
 \end{array}\right)_{x}=U\left(\begin{array}{c}
 \phi_{1}\\
\phi_{2}\\
 \end{array}\right),~U=\left(\begin{array}{cc}
 0& 1\\
-\lambda-u&0\\
 \end{array}\right).
  \end{equation}
The adjoint representation of (\ref{eqns:matrixform}) reads
\begin{equation}
\label{eqns:adjointr1} V_{x}=[U,V].
\end{equation}
 Set
\begin{equation}
\label{eqns:V} V=\sum\limits_{i=1}^{\infty}\left(\begin{array}{cc}
a_{i}& b_{i}\\
c_{i}&-a_{i}\\
 \end{array}\right)\lambda^{-i}.
 \end{equation}
Solving (\ref{eqns:adjointr1}) yields
$$
a_{k}=-\frac{1}{2}b_{k,x},~b_{k+1}=Lb_{k}=-\frac{1}{2}L^{k-1}u,~c_{k}=-\frac{1}{2}
b_{k,xx}-b_{k+1}-b_{k}u,
$$
\begin{equation}
\label{eqns:value}
a_{0}=b_{0}=0,~c_{0}=-1,~a_{1}=0,~b_{1}=1,~c_{1}=-\frac{1}{2}u,~a_{2}=\frac{1}{4}u_{x},
\end{equation}
$$b_{2}=
-\frac{1}{2}u,~c_{2}=\frac{1}{8}(u_{xx}+u^{2}),~b_{3}=\frac{1}{8}(u_{xx}+3u^{2}),\cdots$$
where
$L=-\frac{1}{4}\partial^{2}-u+\frac{1}{2}\partial^{-1}u_{x},~\partial=\frac{\partial}{\partial
x}.$

Set
\begin{equation}
\label{eqns:V(n)}
V^{(n)}=\sum\limits_{i=1}^{n}\left(\begin{array}{cc}
a_{i}& b_{i}\\
c_{i}&-a_{i}\\
 \end{array}\right)\lambda^{n-i}+\left(\begin{array}{cc}
0& 0\\
b_{n+1}&0\\
 \end{array}\right),
 \end{equation}
and take
\begin{equation}
\label{eqns:adjointr2}
\left(\begin{array}{c}
 \phi_{1}\\
\phi_{2}\\
 \end{array}\right)_{t_{n}}=V^{(n)}(u,\lambda)\left(\begin{array}{c}
 \phi_{1}\\
\phi_{2}\\
 \end{array}\right).
 \end{equation}
Then the compatibility of the Eqs.(\ref{eqns:matrixform}) and
(\ref{eqns:adjointr2}) gives rise to the KdV hierarchy
\begin{equation}
\label{eqns:kdvh}
 u_{t_{n}}= -2b_{n+1,x}\equiv \partial \frac{\delta
H_{n}}{\delta u},~n=0,1,\cdots,
\end{equation}
where $H_{n}=4b_{n+2}/2n+1$. We have
\begin{equation}
\label{eqns:bianfen} \frac{\delta\lambda}{\delta
u}=\phi_{1}^{2},~L\phi_{1}^{2}=\lambda\phi_{1}^{2}.
\end{equation}

The higher-order constrained flows of the KdV hierarchy is given by
\cite{zeng2},
\begin{subequations}
\label{eqns:higherorder}
  \begin{align}
    &\frac{\delta H_{n}}{\delta u}-\alpha
\sum\limits_{j=1}^{N}\frac{\delta\lambda_{j}}{\delta u}\equiv
-2b_{n+1}-\alpha\sum\limits_{j=1}^{N}\phi_{1j}^{2}=0,
\\
    &\phi_{1j,x}=\phi_{2j},~\phi_{2j,x}=-(\lambda_{j}+u)\phi_{1j},~j=1,2,\cdots,N.
  \end{align}
\end{subequations}
According to (\ref{eqns:value}), (\ref{eqns:bianfen}) and
(\ref{eqns:higherorder}), we find the Lax representation
(\ref{eqns:adjointr}) for (\ref{eqns:higherorder}) with
\begin{equation}
\label{eqns:18}
 N^{(n)}=\sum\limits_{k=0}^{n}\left(\begin{array}{cc}
a_{k}& b_{k}\\
c_{k}&-a_{k}\\
 \end{array}\right)\lambda^{n-k}+\frac{\alpha}{2}\sum\limits_{j=1}^{N}\frac{1}{\lambda-\lambda_{j}}
 \left(\begin{array}{cc}
 \phi_{1j} \phi_{2j}& - \phi_{1j}^{2}\\
 \phi_{2j}^{2}&-\phi_{1j} \phi_{2j}\\
 \end{array}\right).
 \end{equation}
By taking the so-called Jacobi-Ostrogradsky coordinates\cite{hh}
$$q_{i}=u^{(i-1)},~i=1,\cdots,n-1,$$
$$p_{i}=\frac{\delta H_{n}}{\delta u^{(i)}}=\sum\limits_{l\geq 0}(-\partial)^{l}\frac{\partial H_{n}}
{\partial u^{(i+l)}}$$ and setting
$$\Phi_{1}=(\phi_{11},\phi_{12},\cdots,\phi_{1N})^{T},~\Phi_{2}=(\phi_{21},\phi_{22},\cdots,\phi_{2N})^{T},
~Q=(\phi_{11},\phi_{12},\cdots,\phi_{1N},q_{1},\cdots,q_{n-1})^{T},$$$$P=
(\phi_{21},\phi_{22},\cdots,\phi_{2N},p_{1},\cdots,p_{n-1})^{T},~
\Lambda=diag(\lambda_{1},\lambda_{2},\cdots,\lambda_{N}).$$
Eq.(\ref{eqns:higherorder}) with $\alpha=\frac{2}{4^{n}}$ can be
transformed into a FDIHS\cite{zeng2}
\begin{equation}
\label{eqns:chs}
 Q_{x}=\frac{\partial H}{\partial P},~P_{x}=-\frac{\partial
H}{\partial Q},
 \end{equation}
with$$H=\sum\limits_{i=1}^{n-1}q_{i,x}p_{i}-H_{n}+\frac{1}{2}\langle
\Phi_{2},\Phi_{2}\rangle+\frac{1}{2}\langle \Lambda
\Phi_{1},\Phi_{1}\rangle+\frac{1}{2} q_{1}\langle
\Phi_{1},\Phi_{1}\rangle,$$ where$\langle\rangle $ denotes the inner
production in $R^{N}$. For example, (\ref{eqns:higherorder}) for
$n=0,~\alpha=-2$ gives rise to the Neumann
system\cite{constrainedflow1}, (\ref{eqns:higherorder}) for
$n=1,~\alpha=1$ leads to the Garnier system
\cite{3,constrainedflow1}. When $n=2$, the
Eq.(\ref{eqns:higherorder}) for $\alpha=\frac{1}{8}$ gives the first
higher-order constrained flow\cite{zeng2}
\begin{subequations}
\label{eqns:kdvhigher}
  \begin{align}
    &u_{xx}+3u^{2}=-\frac{1}{2}\sum\limits_{j=1}^{N}
\phi_{1j}^{2}=-\frac{1}{2}\langle
 \Phi_{1},\Phi_{1}\rangle,
\\
    &\phi_{1j,x}=\phi_{2j},~\phi_{2j,x}=-(\lambda_{j}+u)\phi_{1j},~j=1,2,\cdots,N.
  \end{align}
\end{subequations}
 Let
$q_{1}=u,~p_{1}=u_{x},$ (\ref{eqns:kdvhigher}) becomes a FDIHS
(\ref{eqns:chs})
 with $$H=\frac{1}{2}\langle
\Phi_{2},\Phi_{2}\rangle+\frac{1}{2}\langle \Lambda
\Phi_{1},\Phi_{1}\rangle+\frac{1}{2} q_{1}\langle
\Phi_{1},\Phi_{1}\rangle+\frac{1}{2}p^{2}_{1}+q^{3}_{1},$$ and has
the Lax representation (\ref{eqns:adjointr}) with the entries of
$N^{(2)}$ given by
$$A(\lambda)=\frac{1}{4}p_{1}+\frac{1}{16}\sum\limits_{j=1}^{N}\frac{\phi_{1j}
\phi_{2j}}{\lambda-\lambda_{j}},~B(\lambda)=\lambda-\frac{1}{2}q_{1}-\frac{1}{16}\sum\limits_{j=1}^{N}
\frac{\phi_{1j}^{2}}{\lambda-\lambda_{j}} ,$$
$$C(\lambda)=\lambda^{2}-\frac{q_{1}}{2}\lambda-\frac{q_{1}^{2}}{4}-\frac{1}{16}\langle
\Phi_{1},\Phi_{1}\rangle+\frac{1}{16}\sum\limits_{j=1}^{N}\frac{\phi_{2j}^{2}}{\lambda-\lambda_{j}}.$$
With respect to the standard Poisson bracket it is found that
$$
\{A(\lambda),A(\mu)\}=\{B(\lambda),B(\mu)\}=0,~\{C(\lambda),C(\mu)\}=\frac{A(\lambda)-A(\mu)}{4},~~~~~
$$
\begin{equation}
\label{eqns:poisson} \{A(\lambda),B(\mu)\}
=\frac{B(\lambda)-B(\mu)}{8(\lambda-\mu)},~ \{A(\lambda),C(\mu)\}=
\frac{C(\lambda)-C(\mu)}{8(\mu-\lambda)}-\frac{B(\lambda)}{8},
\end{equation}
$$\{B(\lambda),C(\mu)\}=\frac{A(\lambda)-A(\mu)}
{4(\lambda-\mu)}.~~~~~~~~~~~~~~~~~~~~~~~~~~~~~~~~~~~~~~~~~~~~~~~~~~~$$
 It follows from (\ref{eqns:poisson}) that
\begin{equation}
\label{eqns:poisson1}
\{A(\lambda)^{2}+B(\lambda)C(\lambda),A(\mu)^{2}+B(\mu)C(\mu)\}=0.
\end{equation}
When $n=3$, (\ref{eqns:higherorder}) for $\alpha=\frac{1}{32}$
yields the second higher-order constrained flow\cite{zeng2}
\begin{subequations}
\label{eqns:second}
  \begin{align}
    &u_{xxxx}+5u_{x}^{2}+10uu_{xx}+10u^{3}=\frac{1}{2}\langle
 \Phi_{1},\Phi_{1}\rangle,
\\
    &\phi_{1j,x}=\phi_{2j},~\phi_{2j,x}=-(\lambda_{j}+u)\phi_{1j},~j=1,2,\cdots,N.
  \end{align}
\end{subequations}
Let $q_{1}=u,~q_{2}=u_{x},~p_{1}=-u_{xxx}-10uu_{x},~p_{2}=u_{xx},$
(\ref{eqns:second}) becomes a FDIHS (\ref{eqns:chs}) with
$$H=\frac{1}{2}\langle
\Phi_{2},\Phi_{2}\rangle+\frac{1}{2}\langle \Lambda
\Phi_{1},\Phi_{1}\rangle+\frac{1}{2} q_{1}\langle
\Phi_{1},\Phi_{1}\rangle+\frac{1}{2}p^{2}_{2}+q_{2}p_{1}+5q_{1}q_{2}^{2}-\frac{5}{2}q_{1}^{4}.$$
We now consider the Rosochatius deformation $\tilde{N^{(2)}}$ of the
Lax matrix $N^{(2)}$ with
$$\tilde{A}(\lambda)=A(\lambda),~\tilde{B}(\lambda)=B(\lambda),~\tilde{C}(\lambda)=C(\lambda)+
\frac{1}{16}\sum\limits_{j=1}^{N}\frac{\mu_{j}}{(\lambda-\lambda_{j})\phi_{1j}^{2}}.$$
It is not difficult to find that
$\tilde{A}(\lambda),~\tilde{B}(\lambda)$ and $\tilde{C}(\lambda)$
keep the relations of the Poisson brackets (\ref{eqns:poisson}) and
(\ref{eqns:poisson1}).

A direct calculation gives
\begin{equation}
\label{eqns:trace}
\tilde{A}^{2}(\lambda)+\tilde{B}(\lambda)\tilde{C}(\lambda)=-\lambda^{3}+P_{0}
+\sum\limits_{j=1}^{N}\frac{P_{j}}{\lambda-\lambda_{j}}-\frac{1}{256}\sum\limits_{j=1}^{N}\frac{\mu_{j}}
{(\lambda-\lambda_{j})^{2}},
\end{equation}
where
$$
P_{0}=\frac{1}{16}(\langle
  \Phi_{2},\Phi_{2}\rangle+\langle
  \Lambda\Phi_{1},\Phi_{1}\rangle+q_{1}\langle
  \Phi_{1},\Phi_{1}\rangle+2q_{1}^{3}+p_{1}^{2}+\sum\limits_{j=1}^{N}\frac{\mu_{j}}{\phi_{1j}^{2}})$$$$
P_{j}=\frac{p_{1}}{32}\phi_{1j}\phi_{2j}+\frac{1}{16}(\lambda_{j}-\frac{q_{1}}{2})(\phi_{2j}^{2}+\frac{\mu_{j}}{
\phi_{1j}^{2}})+\frac{1}{16}(\lambda_{j}^{2}+\frac{q_{1}}{2}\lambda_{j}+\frac{1}{16}
\langle \Phi_{1},\Phi_{1}\rangle+\frac{q_{1}^{2}}{4})\phi_{1j}^{2}
$$
\begin{equation}
\label{eqns:integral} +\frac{1}{256}\sum\limits_{k\neq
j}\frac{1}{\lambda_{j}-\lambda_{k}}[2\phi_{1j}\phi_{1k}\phi_{2j}\phi_{2k}-\phi_{1j}^{2}(\phi_{2k}^{2}+
\frac{\mu_{k}}{\phi_{1k}^{2}})-\phi_{1k}^{2}(\phi_{2j}^{2}+
\frac{\mu_{j}}{\phi_{1j}^{2}})],~j=1,\cdots,N.
\end{equation}
Choosing $ 8P_{0}=\tilde{H} $ as a Hamiltonian function,  we get the
following Hamiltonian system
\begin{subequations}
\label{eqns:newh}
  \begin{align}
    &q_{1x}=p_{1},~p_{1x}=-\frac{1}{2}\langle
 \Phi_{1},\Phi_{1}\rangle-3q_{1}^{2},
\\
    &\phi_{1jx}=\phi_{2j},~\phi_{2jx}=-\lambda_{j}\phi_{1j}-q_{1}\phi_{1j}+\frac{\mu_{j}}{\phi_{1j}^{3}},
  \end{align}
\end{subequations}
which is the Rosochatius deformation of the first higher-order
constrained flow (\ref{eqns:kdvhigher}).\\
From (\ref{eqns:adjointr}), we have
\begin{equation}
\label{eqns:trace}
\frac{d}{dx}tr\tilde{(N^{(2)}}(\lambda))^{2}=\frac{d}{dx}[\tilde{A}^{2}(\lambda)+\tilde{B}
(\lambda)\tilde{C}(\lambda)]=tr[U,\tilde{(N^{(2)}}(\lambda))^{2}]=0,
\end{equation}
 which implies that
$P_{0},~P_{1},\cdots,P_{N}$ are $N+1$ independent first integrals of
the Hamiltonian system (\ref{eqns:newh}). The equality
(\ref{eqns:poisson1}) of Poisson bracket for
$\tilde{A}(\lambda),~\tilde{B}(\lambda)$ and $\tilde{C}(\lambda)$
indicates that $\{P_{i},P_{j}\}=0,~i,j=0,1,\cdots,N.$ So the
Rosochatius deformation (\ref{eqns:newh}) of the first higher-order
constrained flow  (\ref{eqns:kdvhigher}) is a FDIHS in the Liouville's sense\cite{36}.\\
 {\bf Remark 1.} {\sl For $N=1,~\lambda_{1}=0$, (\ref{eqns:newh}) yields
 $$q_{1xx}=-\frac{1}{2}\phi_{1}^{2}-3q_{1}^{2},$$$$\phi_{1xx}=-q_{1}\phi_{1}+\frac{\mu_{1}}{\phi_{1}^{3}},$$
 which is the well-known generalized integrable H$\acute{e}$non-Heiles system\cite{h1}-\cite{h3}. In fact (\ref{eqns:newh})
 can be regarded as the integrable multidimensional extension of H$\acute{e}$non-Heiles system.}

Similarly, choosing
$$\tilde{H}=\frac{1}{2}\langle
\Phi_{2},\Phi_{2}\rangle+\frac{1}{2}\langle \Lambda
\Phi_{1},\Phi_{1}\rangle+\frac{1}{2} q_{1}\langle
\Phi_{1},\Phi_{1}\rangle+\frac{1}{2}p^{2}_{2}+q_{2}p_{1}+5q_{1}q_{2}^{2}-\frac{5}{2}q_{1}^{4}+\frac{1}{2}
\sum\limits_{j=1}^{N}\frac{\mu_{j}}{\phi_{1j}^{2}},$$ we get the
Rosochatius deformation of the second higher-order constrained flow
(\ref{eqns:second})
\begin{subequations}
\label{eqns:newsecondh}
  \begin{align}
    &q_{1x}=q_{2},~q_{2x}=p_{2},~p_{1x}=10q_{1}^{3}-5q_{2}^{2}-\frac{1}{2}\langle
 \Phi_{1},\Phi_{1}\rangle,~p_{2x}=-10q_{1}q_{2}-p_{1},
\\
    &\phi_{1jx}=\phi_{2j},~\phi_{2jx}=-\lambda_{j}\phi_{1j}-q_{1}\phi_{1j}+\frac{\mu_{j}}{\phi_{1j}^{3}},
  \end{align}
\end{subequations}
which, in the same way,  can be shown to be a FDIHS.\\
In general,  the integrable Rosochatius deformation of the
higher-order constrained flow (\ref{eqns:chs}) is generated by the
following Hamiltonian function
\begin{equation}
\label{eqns:gh}
\tilde{H}=\sum\limits_{i=1}^{n-1}q_{i,x}p_{i}-H_{n}+\frac{1}{2}\langle
\Phi_{2},\Phi_{2}\rangle+\frac{1}{2}\langle \Lambda
\Phi_{1},\Phi_{1}\rangle+\frac{1}{2} q_{1}\langle
\Phi_{1},\Phi_{1}\rangle+\frac{1}{2}\sum\limits_{j=1}^{N}\frac{\mu_{j}}{\phi_{1j}^{2}}.
 \end{equation}

The KdV hierarchy with self-consistent sources is defined
by\cite{constrainedflow5,constrainedflow6,constrainedflow8},\cite{constrainedflow14}-\cite{vk2}
\begin{subequations}
\label{eqns:kdvscs}
  \begin{align}
    &u_{t_{n}}=\partial[\frac{\delta H_{n}}{\delta u}-\alpha \sum\limits_{j=1}^{N}\frac{\delta\lambda_{j}}{\delta u}]\equiv
\partial[-2b_{n+1}-\alpha\sum\limits_{j=1}^{N}\phi_{1j}^{2}],
\\
    &\phi_{1jx}=\phi_{2j},~\phi_{2j,x}=-(\lambda_{j}+u)\phi_{1j},~j=1,2,\cdots,N.
  \end{align}
\end{subequations}
 Since the higher-order constrained flows (\ref{eqns:higherorder}) are just
the stationary equation of the KdV hierarchy with self-consistent
sources (\ref{eqns:kdvscs}), it is obviously that the zero-curvature
representation for the KdV hierarchy with self-consistent sources
(\ref{eqns:kdvscs}) is given by (\ref{eqns:zeror})
with\cite{constrainedflow8}
$$
N^{(n)}=\sum\limits_{i=1}^{n}\left(\begin{array}{cc}
a_{i} & b_{i}\\
 c_{i}&-a_{i}\\
 \end{array}\right)\lambda^{n-i}+\left(\begin{array}{cc}
 0 & 0\\
 b_{n+1}+\frac{\alpha}{2}\sum\limits_{j=1}^{N}
\phi_{1j}^{2}&0\\
 \end{array}\right)$$
 \begin{equation}
\label{eqns:N(n)}+\frac{\alpha}{2}\sum\limits_{j=1}^{N}\frac{1}{\lambda-\lambda_{j}}
 \left(\begin{array}{cc}
 \phi_{1j} \phi_{2j}& - \phi_{1j}^{2}\\
 \phi_{2j}^{2}&-\phi_{1j} \phi_{2j}\\
 \end{array}\right).
 \end{equation}
Eq.(\ref{eqns:kdvscs}) for $n=2,~\alpha=\frac{1}{8}$ gives rise to
the KdV equation with self-consistent sources
\cite{constrainedflow8,constrainedflow14}
\begin{subequations}
\label{eqns:kdvescs2}
  \begin{align}
    &u_{t}=-\frac{1}{4}(u_{xxx}+6uu_{x})-\frac{1}{8}\sum\limits_{j=1}^{N}(\phi_{1j}^{2})_{x},
\\
    &\phi_{1jx}=\phi_{2j},~\phi_{2jx}=-(\lambda_{j}+u)\phi_{1j},
    ~j=1,2,\cdots,N.
  \end{align}
\end{subequations}
 Based on (\ref{eqns:newh}), the  Rosochatius deformation of   KdV equation with
self-consistent sources is given by
\begin{subequations}
\label{eqns:rdkdvscs}
  \begin{align}
    &u_{t}=-\frac{1}{4}(u_{xxx}+6uu_{x})-\frac{1}{8}\sum\limits_{j=1}^{N}(\phi_{1j}^{2})_{x},
\\
    &\phi_{1jx}=\phi_{2j},~\phi_{2jx}=-(\lambda_{j}+u)\phi_{1j}+\frac{\mu_{j}}{\phi_{1j}^{3}},~j=1,\cdots,N
  \end{align}
\end{subequations}
which has the zero-curvature representation (\ref{eqns:zeror}) with
the $N^{(2)}$ given by
$$
  N^{(2)}=\left(\begin{array}{cc}
 \frac{u_{x}}{4}&
 \lambda-\frac{u}{2}\\
-\lambda^{2}-\frac{u}{2}\lambda+\frac{1}{4}u_{xx}+\frac{1}{2}u^{2}+\frac{1}{16}\sum\limits_{j=1}^{N}
\phi_{1j}^{2}&-\frac{u_{x}}{4}
\\
 \end{array}\right)$$
 \begin{equation}
\label{eqns:laxkdvscs}+\frac{1}{16}\sum\limits_{j=1}^{N}\frac{1}{\lambda-\lambda_{j}}\left(\begin{array}{cc}
 \phi_{1j} \phi_{2j}
 & -\phi_{1j}^{2}\\
\phi_{2j}^{2}+\frac{\mu_{j}}{\phi_{1j}^{2}}&-\phi_{1j} \phi_{2j}\\
 \end{array}\right).
\end{equation}
{\bf Remark 2: }{\sl The fact that the stationary equation of
(\ref{eqns:rdkdvscs}) is a FDIHS (\ref{eqns:newh}) and
(\ref{eqns:rdkdvscs}) has zero-curvature representation
(\ref{eqns:zeror}) implies the integrability of the Rosochatius
deformation of the KdV equation with self-consistent source
(\ref{eqns:rdkdvscs}). }

In general the Rosochatius deformation of the KdV hierarchy with
self-consistent sources is given by
\begin{subequations}
\label{eqns:kdvhscs}
  \begin{align}
    &u_{t_{n}}=\partial[\frac{\delta H_{n}}{\delta u}-\frac{2}{4^{n}} \sum\limits_{j=1}^{N}
    \phi_{1j}^{2}],
\\
    &\phi_{1jx}=\phi_{2j},~\phi_{2j,x}=-(\lambda_{j}+u)\phi_{1j}+\frac{\mu_{j}}{\phi_{1j}^{3}},~j=1,2,\cdots,N.
  \end{align}
\end{subequations}
which has the zero-curvature representation (\ref{eqns:zeror}) with
$N^{(n)}$ given by
$$N^{(n)}=\sum\limits_{i=1}^{n}\left(\begin{array}{cc}
a_{i} & b_{i}\\
 c_{i}&-a_{i}\\
 \end{array}\right)\lambda^{n-i}+\left(\begin{array}{cc}
 0&
 0\\
b_{n+1}+\frac{1}{4^{n}}\sum\limits_{j=1}^{N} \phi_{1j}^{2}&0
\\
 \end{array}\right)$$$$+\frac{1}{4^{n}}\sum\limits_{j=1}^{N}\frac{1}{\lambda-\lambda_{j}}\left(\begin{array}{cc}
 \phi_{1j} \phi_{2j}
 & -\phi_{1j}^{2}\\
\phi_{2j}^{2}+\frac{\mu_{j}}{\phi_{1j}^{2}}&-\phi_{1j} \phi_{2j}\\
 \end{array}\right).$$

\section{\bf The  Rosochatius deformed AKNS hierarchy with self-consistent sources}
 For AKNS eigenvalue problem\cite{constrainedflow13}
$$\left(\begin{array}{c}
 \phi_{1}\\
\phi_{2}
\\
 \end{array}\right)_{x}=U\left(\begin{array}{c}
 \phi_{1}\\
\phi_{2}
\\
 \end{array}\right),~U=\left(\begin{array}{cc}
 -\lambda&q\\
r&\lambda
\\
 \end{array}\right),$$
and evolution equation of eigenfunction
$$\left(\begin{array}{c}
 \phi_{1}\\
\phi_{2}
\\
 \end{array}\right)_{t_{n}}=V^{(n)}\left(\begin{array}{c}
 \phi_{1}\\
\phi_{2}
\\
 \end{array}\right),~V^{(n)}=\sum\limits_{i=1}^{n}\left(\begin{array}{cc}
 a_{i}&b_{i}\\
c_{i}&-a_{i}
\\
 \end{array}\right)\lambda^{n-i},$$
 the associated AKNS hierarchy reads
$$u_{t_{n}}=\left(\begin{array}{c}
 q\\
r
\\
 \end{array}\right)_{t_{n}}=J\left(\begin{array}{c}
c_{n+1}\\
b_{n+1}
\\
 \end{array}\right)=J\frac{\delta H_{n+1}}{\delta u}$$
where$$a_{0}=-1,~b_{0}=c_{0}=0,~a_{1}=0,~b_{1}=q,~c_{1}=r,\cdots,$$
$$\left(\begin{array}{c}
c_{n+1}\\
b_{n+1}
\\
 \end{array}\right)=L^{n}\left(\begin{array}{c}
 r\\
q
\\
 \end{array}\right),~L=\frac{1}{2}\left(\begin{array}{cc}
\partial-2r\partial^{-1}q & 2r\partial^{-1}r\\
-2q\partial^{-1}q & -\partial+2q\partial^{-1}r
\\
 \end{array}\right),$$
$$a_{n,x}=qc_{n}-rb_{n},~H_{n}=\frac{2}{n}a_{n+1},~J=\left(\begin{array}{cc}
0 & -2\\
2 & 0
\\
 \end{array}\right),$$
we have
$$\frac{\delta \lambda}{\delta q}=\frac{1}{2}\phi_{2}^{2},~\frac{\delta \lambda}{\delta r}=-
\frac{1}{2}\phi_{1}^{2}.$$ The higher-order constrained flows of the
AKNS hierarchy is given by\cite{constrainedflow6,zeng3}
\begin{subequations}
\label{eqns:AKNS0}
  \begin{align}
    &\frac{\delta H_{n+1}}{\delta u}-\frac{1}{2}\sum\limits_{j=1}^{N}\frac{\delta \lambda_{j}}{\delta u}=
    \left(\begin{array}{c}
c_{n+1}\\
b_{n+1}
\\
 \end{array}\right)-\frac{1}{4}\left(\begin{array}{c}
\langle
\Phi_{2}, \Phi_{2} \rangle\\
-\langle \Phi_{1}, \Phi_{1} \rangle
\\
 \end{array}\right)=0,
\\
    &~\phi_{1jx}=-\lambda_{j}\phi_{1j} +q
\phi_{2j},~\phi_{2jx}=\lambda_{j}\phi_{2j} +r \phi_{1j},
    ~j=1,2,\cdots,N.
  \end{align}
\end{subequations}
(\ref{eqns:AKNS0}) for $n=2$ gives the first higher-order
constrained flow
\begin{subequations}
\label{eqns:AKNS1}
  \begin{align}
    &-q_{xx}+2q^{2}r-\sum\limits_{j=1}^{N}\phi_{1j}^{2}=0,~r_{xx}-2qr^{2}-\sum\limits_{j=1}^{N}\phi_{2j}^{2}=0,
\\
    &~\phi_{1jx}=-\lambda_{j}\phi_{1j} +q
\phi_{2j},~\phi_{2jx}=\lambda_{j}\phi_{2j} +r \phi_{1j},
    ~j=1,2,\cdots,N.
  \end{align}
\end{subequations}

Let
$q_{1}=q,~q_{2}=r,~p_{1}=-\frac{1}{2}r_{x},~p_{2}=-\frac{1}{2}q_{x},~Q=(\phi_{11},\phi_{12},
\cdots,\phi_{1N},q_{1},q_{2})^{T},~P=
(\phi_{21},\phi_{22},\cdots,\phi_{2N},p_{1},p_{2})^{T}.$
(\ref{eqns:AKNS1}) becomes a FDIHS (\ref{eqns:chs}) with
$$ H=-\langle \Lambda
\Phi_{1},\Phi_{2}\rangle+\frac{q_{1}}{2}\langle
\Phi_{2},\Phi_{2}\rangle- \frac{q_{2}}{2}\langle
\Phi_{1},\Phi_{1}\rangle+\frac{1}{2}q_{1}^{2}q_{2}^{2}-2p_{1}p_{2},
$$
 and has the Lax representation
(\ref{eqns:adjointr}) with the entries of Lax matrix $N^{(2)}$ given
by
$$ A(\lambda)=-2\lambda^{2}+q_{1}q_{2}+\frac{1}{2}\sum\limits_{j=1}^{N}\frac{\phi_{1j}\phi_{2j}}
{\lambda-\lambda_{j}},~B(\lambda)=2\lambda
q_{1}+2p_{2}-\frac{1}{2}\sum\limits_{j=1}^{N}\frac{\phi_{1j}^{2}}
{\lambda-\lambda_{j}},$$
$$ C(\lambda)=2\lambda q_{2}-2p_{1}+\frac{1}{2}\sum\limits_{j=1}^{N}\frac{\phi_{2j}^{2}}
{\lambda-\lambda_{j}}.$$ With respect to the standard Poisson
bracket it is found that
$$\{A(\lambda),A(\mu)\}=\{B(\lambda),B(\mu)\}=\{C(\lambda),C(\mu)\}=0,~\{A(\lambda),B(\mu)\}
=\frac{B(\lambda)-B(\mu)}{\lambda-\mu},$$
\begin{equation}
\label{eqns:AKNS3}
\{A(\lambda),C(\mu)\}=\frac{C(\mu)-C(\lambda)}{\lambda-\mu},
~\{B(\lambda),C(\mu)\}=\frac{2[A(\lambda)-A(\mu)]} {\lambda-\mu},
\end{equation}
which gives rise to (\ref{eqns:poisson1}).

Now we consider the  Rosochatius deformation $\tilde{N}^{(2)}$ of
the  Lax matrix $N^{(2)}$
\begin{equation}
\label{eqns:AKNS5}
\tilde{A}(\lambda)=A(\lambda),~\tilde{B}(\lambda)=B(\lambda),~\tilde{C}(\lambda)=C(\lambda)+
\frac{1}{2}\sum\limits_{j=1}^{N}\frac{\mu_{j}}{(\lambda-\lambda_{j})\phi_{1j}^{2}}.
 \end{equation}
It is easy to find that the elements in $\tilde{N^{(2)}}$ still keep
the relations of the Poisson brackets (\ref{eqns:AKNS3}) and
(\ref{eqns:poisson1}).

A direct calculation gives
\begin{equation}
\label{eqns:AKNS6}
\tilde{A}^{2}(\lambda)+\tilde{B}(\lambda)\tilde{C}(\lambda)=4\lambda^{4}+P_{0}\lambda+P_{1}+\sum\limits_{j=1}^{N}\frac{P_{j}}{\lambda-\lambda_{j}}-\frac{1}{4}\sum\limits_{j=1}^{N}
 \frac{\mu_{j}}{(\lambda-\lambda_{j})^{2}}
 \end{equation}
where
$$P_{0}=-4q_{1}p_{1}+4q_{2}p_{2}-2\langle\Phi_{1},
 \Phi_{2}\rangle,$$
$$P_{1}=-4p_{1}p_{2}+q_{1}^{2}q_{2}^{2}+q_{1}(\langle\Phi_{2},\Phi_{2}\rangle+
 \sum\limits_{j=1}^{N}\frac{\mu_{j}}{\phi_{1j}^{2}})
-q_{2}\langle\Phi_{1},\Phi_{1}\rangle-2
 \langle
 \Lambda\Phi_{1},\Phi_{2}\rangle,$$
\begin{equation}
\label{eqns:AKNS7}
P_{j+1}=(\lambda_{j}q_{1}+p_{2})(\phi_{2j}^{2}+\frac{\mu_{j}}{\phi_{1j}^{2}})
-(\lambda_{j}q_{2}-p_{1})\phi_{1j}^{2}+(q_{1}q_{2}-2\lambda_{j}^{2})
\phi_{1j}\phi_{2j}~~~~~~~~
\end{equation}
$$
+\frac{1}{4}\sum\limits_{k\neq
j}\frac{1}{\lambda_{j}-\lambda_{k}}[2\phi_{1j}\phi_{2j}\phi_{1k}\phi_{2k}-\phi_{1j}^{2}
(\phi_{2k}^{2}+\frac{\mu_{k}}{\phi_{1k}^{2}})-\phi_{1k}^{2}(\phi_{2j}^{2}+\frac{\mu_{j}}{\phi_{1j}^{2}})],~
j=1,\cdots,N.
$$
Choosing $\frac{1}{2}P_{1}=\tilde{H}$ as a Hamiltonian function, we
get the following Hamiltonian system
\begin{subequations}
\label{eqns:AKNS8}
  \begin{align}
    &q_{1x}=-2p_{2},~q_{2x}=-2p_{1},\\
    &p_{1x}=-\frac{1}{2}\langle \Phi_{2},
\Phi_{2}
\rangle-\sum\limits_{j=1}^{N}\frac{\mu_{j}}{2\phi_{1j}^{3}}-
q_{2}^{2}q_{1},~p_{2x}=\frac{1}{2}\langle \Phi_{1}, \Phi_{1}
\rangle-q_{1}^{2}q_{2},
\\
    &\phi_{1jx}=-\lambda_{j}\phi_{1j} +q_{1}
\phi_{2j},~\phi_{2jx}=\lambda_{j}\phi_{2j} +q_{2} \phi_{1j}+
\frac{\mu_{j} q_{1}}{\phi_{1j}^{3}},~j=1,\cdots,N
  \end{align}
\end{subequations}
which  has the Lax representation (\ref{eqns:adjointr}) with
$N^{(2)}$ replaced by $\tilde{N^{(2)}}$. Then (\ref{eqns:poisson1})
and (\ref{eqns:trace}) implies that $P_{0},~P_{1},~\cdots,P_{N+1}$
are $N+2$ independent first integrals in involution, so
(\ref{eqns:AKNS8}) is a FDIHS\cite{36}.

In general, in the similar way as in Sec.2, (\ref{eqns:AKNS0})
becomes a FDIHS (\ref{eqns:chs}) with
$$H=\sum\limits_{i=1}^{n}q_{i,x}p_{i}-H_{n+1}-\langle \Lambda
\Phi_{1},\Phi_{2}\rangle+\frac{q_{1}}{2}\langle
\Phi_{2},\Phi_{2}\rangle- \frac{q_{2}}{2}\langle
\Phi_{1},\Phi_{1}\rangle.$$ Then the integrable Rosochatius
deformation of the higher-order constrained flow (\ref{eqns:AKNS0})
is generated by the following Hamiltonian function
\begin{equation}
\label{eqns:gh}
\tilde{H}=\sum\limits_{i=1}^{n}q_{i,x}p_{i}-H_{n+1}-\langle \Lambda
\Phi_{1},\Phi_{2}\rangle+\frac{q_{1}}{2}\langle
\Phi_{2},\Phi_{2}\rangle- \frac{q_{2}}{2}\langle
\Phi_{1},\Phi_{1}\rangle+\frac{1}{2}\sum\limits_{j=1}^{N}\frac{\mu_{j}q_{1}}{\phi_{1j}^{2}}.
 \end{equation}

The AKNS hierarchy with self-consistent sources
is\cite{constrainedflow6}
\begin{subequations}
\label{eqns:AKNS12}
  \begin{align}
    &\left(\begin{array}{c}
q\\
r
\\
 \end{array}\right)_{t_{n}}=J[\frac{\delta H_{n+1}}{\delta u}-\frac{1}{2}\sum\limits_{j=1}^{N}\frac{\delta \lambda_{j}}
 {\delta u}]=J
    [\left(\begin{array}{c}
c_{n+1}\\
b_{n+1}
\\
 \end{array}\right)-\frac{1}{4}\left(\begin{array}{c}
\langle
\Phi_{2}, \Phi_{2} \rangle\\
-\langle \Phi_{1}, \Phi_{1} \rangle
\\
 \end{array}\right)],
\\
    &~\phi_{1jx}=-\lambda_{j}\phi_{1j} +q
\phi_{2j},~\phi_{2jx}=\lambda_{j}\phi_{2j} +r \phi_{1j},
    ~j=1,2,\cdots,N.
  \end{align}
\end{subequations}
When $n=2$, the AKNS equation with self-consistent sources
reads\cite{constrainedflow6}
\begin{subequations}
\label{eqns:AKNS4}
  \begin{align}
    &2q_{t}=-q_{xx}+2q^{2}r-\sum\limits_{j=1}^{N}\phi_{1j}^{2},~2r_{t}=r_{xx}-2qr^{2}-
\sum\limits_{j=1}^{N}\phi_{2j}^{2},
\\
    &\phi_{1jx}=-\lambda_{j}\phi_{1j} +q
\phi_{2j},~\phi_{2jx}=\lambda_{j}\phi_{2j} +r
\phi_{1j},~j=1,\cdots,N.
  \end{align}
\end{subequations}
Based on (\ref{eqns:AKNS8}), we
 obtain the integrable Rosochatius deformation of the  AKNS equation with
self-consistent sources
\begin{subequations}
\label{eqns:AKNS9}
  \begin{align}
    &2q_{t}=-q_{xx}+2q^{2}r-\sum\limits_{j=1}^{N}\phi_{1j}^{2},~2r_{t}=r_{xx}-2qr^{2}-
\sum\limits_{j=1}^{N}\phi_{2j}^{2}-\sum\limits_{j=1}^{N}\frac{\mu_{j}}{\phi_{1j}^{2}},
\\
    &\phi_{1jx}=-\lambda_{j}\phi_{1j} +q
\phi_{2j},~\phi_{2jx}=\lambda_{j}\phi_{2j} +r \phi_{1j}+
\frac{\mu_{j} q}{\phi_{1j}^{3}},~~~~j=1,\cdots,N
  \end{align}
\end{subequations}
 which has
the zero-curvature representation (\ref{eqns:zeror})  with $N^{(2)}$
\begin{equation}
\label{eqns:AKNS10} N^{(2)}=\left(\begin{array}{cc}
 -2\lambda^{2}+qr& 2\lambda q-q_{x}\\
2\lambda r+r_{x} &2\lambda^{2}-qr\\
 \end{array}\right)+\frac{1}{2}\sum\limits_{j=1}^{N}\frac{1}{\lambda-\lambda_{j}}
 \left(\begin{array}{cc}
 \phi_{1j} \phi_{2j}& - \phi_{1j}^{2}\\
 \phi_{2j}^{2}+\frac{\mu_{j}}{\phi_{1j}^{2}}&-\phi_{1j} \phi_{2j}\\
 \end{array}\right).
\end{equation}

In general the integrable Rosochatius deformation of the AKNS
hierarchy with self-consistent sources is given by
\begin{subequations}
\label{eqns:AKNS13}
  \begin{align}
    &\left(\begin{array}{c}
q\\
r
\\
 \end{array}\right)_{t_{n}}=J
    [\left(\begin{array}{c}
c_{n+1}\\
b_{n+1}
\\
 \end{array}\right)-\frac{1}{4}\left(\begin{array}{c}
\langle
\Phi_{2}, \Phi_{2} \rangle+\sum\limits_{j=1}^{N}\frac{\mu_{j}}{\phi_{1j}^{2}}\\
-\langle \Phi_{1}, \Phi_{1} \rangle
\\
 \end{array}\right)],
\\
    &~\phi_{1jx}=-\lambda_{j}\phi_{1j} +q
\phi_{2j},~\phi_{2jx}=\lambda_{j}\phi_{2j} +r
\phi_{1j}+\frac{\mu_{j} q}{\phi_{1j}^{3}},
    ~j=1,2,\cdots,N.
  \end{align}
\end{subequations}
which has the zero-curvature representation (\ref{eqns:zeror}) with
$N^{(n)}$ given by
$$N^{(n)}=V^{(n)}+\frac{1}{2}\sum\limits_{j=1}^{N}\frac{1}{\lambda-\lambda_{j}}\left(\begin{array}{cc}
 \phi_{1j} \phi_{2j}
 & -\phi_{1j}^{2}\\
\phi_{2j}^{2}+\frac{\mu_{j}}{\phi_{1j}^{2}}&-\phi_{1j} \phi_{2j}\\
 \end{array}\right).$$
{\bf Remark 3.} {\sl In contrast with the Rosochatius deformation of
KdV hierarchy with self-consistent sources, the Rosochatius
deformation of AKNS hierarchy with self-consistent sources has the
deformed term in both (48a) and (48b).}

\section{\bf The Rosochatius deformed  mKdV hierarchy with self-consistent sources}
For mKdV spectral problem\cite{constrainedflow11}
$$\left(\begin{array}{c}
 \phi_{1}\\
\phi_{2}
\\
 \end{array}\right)_{x}=U\left(\begin{array}{c}
 \phi_{1}\\
\phi_{2}
\\
 \end{array}\right),~U=\left(\begin{array}{cc}
 -u&\lambda\\
\lambda&u
\\
 \end{array}\right),$$
and evolution equation of eigenfunction
$$\left(\begin{array}{c}
 \phi_{1}\\
\phi_{2}
\\
 \end{array}\right)_{t_{n}}=V^{(n)}\left(\begin{array}{c}
 \phi_{1}\\
\phi_{2}
\\
 \end{array}\right),~V^{(n)}=\sum\limits_{i=1}^{n-1}\left(\begin{array}{cc}
a_{i}\lambda&b_{i}\\
c_{i}&-a_{i}\lambda
\\
 \end{array}\right)\lambda^{2n-2i-3}+\left(\begin{array}{cc}
a_{n}&0\\
0&-a_{n}
\\
 \end{array}\right),$$
 the associated mKdV
hierarchy reads
$$u_{t_{n}}=-\partial a_{n}=\partial\frac{\delta H_{n}}{\delta u},$$
where$$a_{0}=0,~b_{0}=c_{0}=1,~a_{1}=-u,~b_{1}=-\frac{u^{2}}{2}+\frac{u_{x}}{2},
~c_{1}=-\frac{u^{2}}{2}-\frac{u_{x}}{2},\cdots,$$
$$a_{n+1}=La_{n},~L=\frac{1}{4}\partial^{2}-u\partial^{-1}u\partial$$
$$b_{n}=\partial^{-1}u\partial a_{n}-\frac{1}{2}a_{nx},~c_{n}=
\partial^{-1}u\partial a_{n}+\frac{1}{2}a_{nx},$$
we have
$$\frac{\delta \lambda}{\delta u}=\frac{1}{2}\phi_{1}\phi_{2}.$$
The higher-order constrained flows of the mKdV hierarchy is
\begin{subequations}
\label{eqns:mkdv0}
  \begin{align}
    &\frac{\delta H_{n}}{\delta u}+2
\sum\limits_{j=1}^{N}\frac{\delta\lambda_{j}}{\delta u}\equiv
-a_{n}+\sum\limits_{j=1}^{N}\phi_{1j}\phi_{2j}=0,
\\
    &\phi_{1j,x}=-u\phi_{1j}+\lambda_{j}\phi_{2j},~\phi_{2j,x}=\lambda_{j}\phi_{1j}+u\phi_{2j},~j=1,2,\cdots,N.
  \end{align}
\end{subequations}
When $n=2$, (\ref{eqns:mkdv0}) gives the first higher-order
constrained flow
\begin{subequations}
\label{eqns:mkdv1}
  \begin{align}
    &u_{xx}-2u^{3}=-4\sum\limits_{j=1}^{N}\phi_{1j}\phi_{2j}=-4\langle
\Phi_{1}, \Phi_{2} \rangle,
\\
    &\phi_{1jx}=\lambda_{j}\phi_{2j} -u\phi_{1j},~\phi_{2jx}=\lambda_{j}\phi_{1j} +u
    \phi_{2j},~j=1,\cdots,N.
  \end{align}
\end{subequations}
Let $q_{1}=u,~p_{1}=-\frac{u_{x}}{4},$ it becomes a FDIHS
(\ref{eqns:chs})
 with
$$
 H=-q_{1}\langle
\Phi_{1},\Phi_{2}\rangle+\frac{1}{2}\langle
\Lambda\Phi_{2},\Phi_{2}\rangle- \frac{1}{2}\langle
\Lambda\Phi_{1},\Phi_{1}\rangle-2p_{1}^{2}+\frac{1}{8} q_{1}^{4},
$$
and has the Lax representation (\ref{eqns:adjointr}) with the
entries of Lax matrix $N^{(2)}$ given by
 \begin{equation}
 \label{eqns:mkdv3}
A(\lambda)=-q_{1}\lambda+\sum\limits_{j=1}^{N}\frac{\lambda\phi_{1j}\phi_{2j}}
{\lambda^{2}-\lambda_{j}^{2}},~B(\lambda)=\lambda^{2}
-\frac{1}{2}q_{1}^{2}-2p_{1}-\sum\limits_{j=1}^{N}\frac{\lambda_{j}\phi_{1j}^{2}}
{\lambda^{2}-\lambda_{j}^{2}},
 \end{equation}
 $$ C(\lambda)=\lambda^{2}-\frac{1}{2}q_{1}^{2} +2p_{1}+\sum\limits_{j=1}^{N}
\frac{\lambda_{j}\phi_{2j}^{2}}
{\lambda^{2}-\lambda_{j}^{2}}.~~~~~~~~~~~~~~~~~~~~~~~~~~~~~~~$$ With
respect to the standard Poisson bracket, a direct calculation gives
$$
\{A(\lambda),A(\mu)\}=\{B(\lambda),B(\mu)\}=\{C(\lambda),C(\mu)\}=0,~\{A(\lambda),B(\mu)\}
=2\lambda\frac{B(\lambda)-B(\mu)}{\lambda^{2}-\mu^{2}},$$
\begin{equation}
\label{eqns:mkdv4}
\{A(\lambda),C(\mu)\}=2\lambda\frac{C(\lambda)-C(\mu)}{\mu^{2}-\lambda^{2}},
~\{B(\lambda),C(\mu)\}=\frac{4[A(\mu)\mu-A(\lambda)\lambda]}
{\mu^{2}-\lambda^{2}}~~~~~~~~~~~~~~
\end{equation}
which leads to (\ref{eqns:poisson1}).\\
Now we consider the integrable Rosochatius deformation
$\tilde{N}^{(2)}$ of the  Lax matrix $N^{(2)}$
\begin{equation}
\label{eqns:AKNS5}
\tilde{A}(\lambda)=A(\lambda),~\tilde{B}(\lambda)=B(\lambda),~\tilde{C}(\lambda)=C(\lambda)+
\sum\limits_{j=1}^{N}\frac{\mu_{j}\lambda_{j}}{(\lambda^{2}-\lambda_{j}^{2})\phi_{1j}^{2}}.
 \end{equation}
It is not difficult to find that the elements in $\tilde{N^{(2)}}$
still keep the relations of the Poisson
brackets (\ref{eqns:mkdv4}) and (\ref{eqns:poisson1}).\\
 A direct calculation gives
\begin{equation}
\label{eqns:mkdv7}
\tilde{A}^{2}(\lambda)+\tilde{B}(\lambda)\tilde{C}(\lambda)=\lambda^{4}+P_{0}+
 \sum\limits_{j=1}^{N}\frac{P_{j}}{\lambda^{2}-\lambda_{j}^{2}}- \sum\limits_{j=1}^{N}
 \frac{\lambda_{j}\mu_{j}}{(\lambda^{2}-\lambda_{j}^{2})^{2}}
 \end{equation}
where $$P_{0}=-2q_{1}\langle \Phi_{1},\Phi_{2}\rangle
 +\langle \Lambda\Phi_{2},\Phi_{2}\rangle-\langle \Lambda\Phi_{1},\Phi_{1}\rangle+\sum
 \limits_{j=1}^{N}\frac{\lambda_{j}\mu_{j}}{\phi_{1j}^{2}}
 +\frac{1}{4}q_{1}^{4}-4p_{1}^{2}$$
\begin{equation}
\label{eqns:mkdv8}
P_{j}=-2q_{1}\lambda_{j}^{2}\phi_{1j}\phi_{2j}+\lambda_{j}^{3}(\phi_{2j}^{2}+\frac{\mu_{j}}
{\phi_{1j}^{2}})
-(\frac{1}{2}q_{1}^{2}+2p_{1})(\lambda_{j}\phi_{2j}^{2}+\frac{\lambda_{j}\mu_{j}}{\phi_{1j}^{2}})
-\lambda_{j}^{3}\phi_{1j}^{2}+(\frac{1}{2}q_{1}^{2}-2p_{1})\lambda_{j}\phi_{1j}^{2}
\end{equation}
$$-\sum\limits_{k\neq
j}\frac{1}{\lambda_{j}^{2}-\lambda_{k}^{2}}[2\lambda_{j}^{2}\phi_{1j}\phi_{2j}\phi_{1k}\phi_{2k}
+\lambda_{j}\lambda_{k}\phi_{1j}^{2}(\phi_{2k}^{2}+\frac{\mu_{k}}{\phi_{1k}^{2}})+
\lambda_{j}\lambda_{k}\phi_{1k}^{2}(\phi_{2j}^{2}+\frac{\mu_{j}}{\phi_{1j}^{2}})].
$$
Choosing $\frac{1}{2}P_{0}=\tilde{H}$ as Hamiltonian function, we
get the following Hamiltonian system
\begin{subequations}
\label{eqns:mkdv9}
  \begin{align}
    &q_{1x}=-4p_{1},~p_{1x}=\langle \Phi_{1}, \Phi_{2} \rangle-\frac{1}{2}q_{1}^{3},
\\
    &\phi_{1jx}=\lambda_{j}\phi_{2j} -q_{1}\phi_{1j},~\phi_{2jx}=\lambda_{j}\phi_{1j} +q_{1}\phi_{2j}+
\frac{\lambda_{j}\mu_{j} }{\phi_{1j}^{3}},~j=1,\cdots,N
  \end{align}
\end{subequations}
 which has the
Lax representation (\ref{eqns:adjointr}) with $N^{(2)}$ replaced by
$\tilde{N}^{(2)}$ and is a FDIHS.

Similarly, (\ref{eqns:mkdv0}) can be transformed into a FDIHS
(\ref{eqns:chs}) with $N+1$ independent first integrals
$P_{0},~P_{1},~\cdots,P_{N}$ in involution
$$H=\sum\limits_{i=1}^{n-1}q_{i,x}p_{i}-H_{n}-q_{1}\langle
\Phi_{1},\Phi_{2}\rangle+\frac{1}{2}\langle
\Lambda\Phi_{2},\Phi_{2}\rangle- \frac{1}{2}\langle
\Lambda\Phi_{1},\Phi_{1}\rangle.$$ Then the integrable Rosochatius
deformation of the higher-order constrained flow (\ref{eqns:mkdv0})
is generated by the following Hamiltonian function
\begin{equation}
\label{eqns:gh}
\tilde{H}=\sum\limits_{i=1}^{n-1}q_{i,x}p_{i}-H_{n}-q_{1}\langle
\Phi_{1},\Phi_{2}\rangle+\frac{1}{2}\langle
\Lambda\Phi_{2},\Phi_{2}\rangle- \frac{1}{2}\langle
\Lambda\Phi_{1},\Phi_{1}\rangle+\frac{1}{2}\sum\limits_{j=1}^{N}\frac{\lambda_{j}\mu_{j}}{\phi_{1j}^{2}}.
 \end{equation}

The mKdV hierarchy with self-consistent sources is
\begin{subequations}
\label{eqns:mkdv15}
  \begin{align}
    &u_{t_{n}}=\partial[\frac{\delta H_{n}}{\delta u}+2
\sum\limits_{j=1}^{N}\frac{\delta\lambda_{j}}{\delta u}]\equiv
\partial[-a_{n}+\sum\limits_{j=1}^{N}\phi_{1j}\phi_{2j}],
\\
    &\phi_{1j,x}=-u\phi_{1j}+\lambda_{j}\phi_{2j},~\phi_{2j,x}=\lambda_{j}\phi_{1j}+u\phi_{2j},~j=1,2,\cdots,N.
  \end{align}
\end{subequations}
When $n=2$, (\ref{eqns:mkdv15}) gives the mKdV equation with
self-consistent sources
\begin{subequations}
\label{eqns:mkdv5}
  \begin{align}
    &u_{t}=\frac{u_{xxx}}{4}-\frac{3}{2}u^{2}u_{x}+\sum\limits_{j=1}^{N}(\phi_{1j}\phi_{2j})_{x},
\\
    &\phi_{1jx}=\lambda_{j}\phi_{2j}
-u\phi_{1j},~\phi_{2jx}=\lambda_{j}\phi_{1j} +u
\phi_{2j},~~~~j=1,\cdots,N.
  \end{align}
\end{subequations}
Based on (\ref{eqns:mkdv9}), the integrable Rosochatius deformation
of the mKdV equation with self-consistent sources is given by
\begin{subequations}
\label{eqns:mkdv10}
  \begin{align}
    &u_{t}=\frac{u_{xx}}{4}-\frac{3}{2}u^{2}u_{x}+\sum\limits_{j=1}^{N}(\phi_{1j}\phi_{2j})_{x},
\\
    &\phi_{1jx}=\lambda_{j}\phi_{2j}
-u\phi_{1j},~\phi_{2jx}=\lambda_{j}\phi_{1j} +u \phi_{2j}+
\frac{\lambda_{j}\mu_{j} }{\phi_{1j}^{3}},~j=1,\cdots,N.
  \end{align}
\end{subequations}
which has the zero-curvature representation (\ref{eqns:zeror})  with
$N^{(2)}$ given by $$ N^{(2)}=\left(\begin{array}{cc}
 -u\lambda^{2}-\frac{u_{xx}}{4}+\frac{u^{3}}{2}& \lambda^{3}-(\frac{u^{2}}{2}-\frac{u_{x}}{2})\lambda\\
\lambda^{3}-(\frac{u^{2}}{2}+\frac{u_{x}}{2})\lambda &u\lambda^{2}+\frac{u_{xx}}{4}-\frac{u^{3}}{2}\\
 \end{array}\right)+\left(\begin{array}{cc}
 -\sum\limits_{j=1}^{N}\phi_{1j}\phi_{2j}& 0\\
0 & \sum\limits_{j=1}^{N}\phi_{1j}\phi_{2j}\\
 \end{array}\right)
$$
\begin{equation}
\label{eqns:mkdv11}
+\sum\limits_{j=1}^{N}\frac{\lambda}{\lambda^{2}-\lambda_{j}^{2}}
 \left(\begin{array}{cc}
 \lambda\phi_{1j} \phi_{2j}& - \lambda_{j}\phi_{1j}^{2}\\
 \lambda_{j}(\phi_{2j}^{2}+\frac{\mu_{j}}{\phi_{1j}^{2}})&-\lambda\phi_{1j} \phi_{2j}\\
 \end{array}\right)
 \end{equation}
In general the integrable  Rosochatius deformation of the mKdV
hierarchy with self-consistent sources is given by
\begin{subequations}
\label{eqns:mkdv13}
  \begin{align}
    &u_{t_{n}}=\partial[\frac{\delta H_{n}}{\delta u}+2\sum\limits_{j=1}^{N}\frac{\delta \lambda_{j}}{\delta u}]
    =\partial[-a_{n}+\sum\limits_{j=1}^{N}\phi_{1j}\phi_{2j}]
\\
    &\phi_{1jx}=\lambda_{j}\phi_{2j}
-u\phi_{1j},~\phi_{2jx}=\lambda_{j}\phi_{1j} +u \phi_{2j}+
\frac{\lambda_{j}\mu_{j} }{\phi_{1j}^{3}},~~~~j=1,\cdots,N.
  \end{align}
\end{subequations}
which has the zero-curvature representation (\ref{eqns:zeror}) with
$N^{(n)}$ given by
$$N^{(n)}=\sum\limits_{i=1}^{n-1}\left(\begin{array}{cc}
a_{i}\lambda&b_{i}\\
c_{i}&-a_{i}\lambda
\\
 \end{array}\right)\lambda^{2n-2i-3}+\left(\begin{array}{cc}
 a_{n}-\sum\limits_{j=1}^{N}\phi_{1j}\phi_{2j}& 0\\
0 & a_{n}+\sum\limits_{j=1}^{N}\phi_{1j}\phi_{2j}\\
 \end{array}\right)$$$$+\sum\limits_{j=1}^{N}\frac{\lambda}{\lambda^{2}-\lambda_{j}^{2}}
 \left(\begin{array}{cc}
 \lambda\phi_{1j} \phi_{2j}& - \lambda_{j}\phi_{1j}^{2}\\
 \lambda_{j}(\phi_{2j}^{2}+\frac{\mu_{j}}{\phi_{1j}^{2}})&-\lambda\phi_{1j} \phi_{2j}\\
 \end{array}\right).$$

\section{Conclusion}
Rosochatius-type integrable systems have important physical
applications. However, the studies on Rosochatius deformation are
limited to few finite dimensional integrable Hamiltonian
system(FDIHS) at present. The main purpose of this paper is to
propose a systematic method to generalize the Rosochatius
deformation for FDIHS to the Rosochatius deformation for infinite
dimensional integrable equations. We first construct infinite number
of integrable  Rosochatius deformations of FDIHSs obtained from
higher-order constrained flows of some soliton hierarchies. Then,
based on the Rosochatius deformed higher-order constrained flows, we
establish the integrable Rosochatius deformations of some soliton
hierarchies with self-consistent sources. The Rosochatius
deformations of the KdV hierarchy with self-consistent sources, of
the AKNS hierarchy with self-consistent sources and of the mKdV
hierarchy with self-consistent sources, together with their Lax
pairs are obtained. The approach presented here can be applied to
other cases.

\section*{Acknowledgments}

This work is supported by National Basic Research Program of China
(973 Program) (2007CB814800) and National Natural Science Foundation
of China (10671121).


\def\cprime{$'$} \def\cprime{$'$} \def\cprime{$'$} \def\cprime{$'$}
  \def\cprime{$'$}

\end{document}